\pdfoutput=1
\documentclass[10pt,conference]{IEEEtran}
\usepackage[T1]{fontenc}
\usepackage[utf8]{inputenc}
\usepackage{calc,amsmath,amssymb,booktabs,array,graphicx,hyperref}
\usepackage{xurl}
\usepackage{xcolor,framed}
\definecolor{shadecolor}{gray}{0.94}
\renewenvironment{quote}{\begin{shaded}\small\noindent}{\end{shaded}}
\providecommand{\tightlist}{\setlength{\itemsep}{0pt}\setlength{\parskip}{0pt}}
\providecommand{\pandocbounded}[1]{#1}

\title{Mismeasuring the Life of a Line of Code: Lessons Learned the Hard Way}
\author{\IEEEauthorblockN{Pavel Gurov}\IEEEauthorblockA{Independent Researcher\\ORCID: \href{https://orcid.org/0009-0004-7407-8617}{0009-0004-7407-8617}}}
\begin{document}
\maketitle
\begin{abstract}
Studies of how long lines of code live, including lines written by AI
agents, rest on choices that look technical: when a line counts as gone,
what to do with lines still alive, whether to pool projects, and how to
evaluate a model. We studied 32.5 million line births in 120 TypeScript
repositories and got several of these choices wrong at first. For ten of
them, we compare the result as we first computed it, or as commonly
done, with the result after correction. All ten change conclusions, or
would have. Six were our own mistakes. Counting moved and rewritten
lines as dead shortens a line's median life from 3.0 years to 1.3.
Counting any change as a death hides an outcome that can be predicted:
within-project concordance falls from 0.71 to 0.56. Regressing observed
lifetimes discards three quarters of the lines. Keeping rewritten lines
at risk makes protective effects grow with age. Lines are born and
removed together, so treating them as independent misleads twice: nine
of fifteen effects look significant only under that assumption, and a
split by file makes a barely predictable outcome look good
(within-project concordance 0.71 instead of 0.61). After the
corrections, removal depends on a line's age and project, but our models
could barely rank it within a project. Local revision they could rank
(0.71). Obvious cases aside, the top 1\% of a project's new lines are
revised at 2.3 times the base rate. Each pitfall leaves a symptom; we
give a checklist.

\end{abstract}
\begin{quote}
\textbf{Relation to an earlier preprint.} This paper is a new study of
the data of arXiv:2606.04993 by the same author. Several results of that
preprint did not survive the checks reported here and should no longer
be cited.
\end{quote}

\section{Introduction}\label{introduction}

Studies of how long code lives have moved from projects and files down
to single lines. Spinellis et al.~{[}1{]} followed billions of line and
token lifetimes across long-lived repositories. Rahman and Shihab
{[}2{]} compared how long lines written by AI agents and by people
survive before they are changed. The answers matter in practice too:
tools that highlight fragile code, prioritise review or estimate how
long a change will last all depend on them.

Lines, however, are awkward subjects. A line can be edited, reformatted,
moved to another file or changed by renaming an identifier. It can be
deleted along with hundreds of others when a feature goes, or stay
untouched for years in a module nobody works on. A study of line
lifetimes therefore has to decide when a line counts as gone and what to
do with lines still alive when the data end. It has to decide whether
projects as different as ours can be treated as one population, and how
to check a model that predicts a line's fate. These decisions rarely
reach a paper's abstract.

We learned their weight the hard way. Some early results of our study
were spectacular, and wrong. In one version of our model, the entropy of
a line's characters was the strongest protective feature, with a p-value
below \(10^{-130}\). Then a reader outside the project wrote to us that
entropy was a red herring, and it turned out to measure the line's
length. A classifier predicting a line's death reached an AUC of 0.92,
which we did not believe; the training and test data shared lines added
by the same commits. And our data loaders quietly did something other
than what our paper said: a line that had already been rewritten, and so
no longer existed in its original form, was still counted as a line that
could be deleted, for three years on average. Each of these errors
changed a conclusion we had already written down.

This paper turns those errors, and a few we avoided, into ten pitfalls.
For each, we show the result as we first computed it, or as it is
commonly computed, and the result after correction. We explain why the
two differ and how the problem could have been noticed earlier. Six
pitfalls are mistakes we made, one we nearly made, and three are
practices we avoided but found in published studies (§6). Fig. 1 shows
the most striking case: the same model looks strong or weak depending
only on how the lines are split between training and test.

\begin{figure}
\centering
\pandocbounded{\includegraphics[keepaspectratio,alt={The same model, evaluated on lines separated ever more strictly between training and test. Lines born in one half-year are split in half by file, commit, directory or repository; forward evaluation trains on the past at five cutoffs. Between a file split and forward evaluation at the matching cutoff, removal (with the file or in bulk) loses 0.11 of concordance, both over all test lines and within projects, and local revision loses 0.06 and 0.05. Markers are means over five draws of each split, or over the five cutoffs, and bars show their range. First sample; the second, with one draw per split, gives the same picture.}]{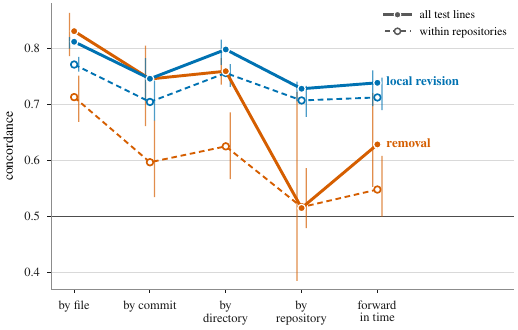}}
\caption{The same model, evaluated on lines separated ever more strictly
between training and test. Lines born in one half-year are split in half
by file, commit, directory or repository; forward evaluation trains on
the past at five cutoffs. Between a file split and forward evaluation at
the matching cutoff, removal (with the file or in bulk) loses 0.11 of
concordance, both over all test lines and within projects, and local
revision loses 0.06 and 0.05. Markers are means over five draws of each
split, or over the five cutoffs, and bars show their range. First
sample; the second, with one draw per split, gives the same picture.}
\end{figure}

Once the pitfalls are avoided, our data tell a simple story. Lines are
removed with the modules and projects around them, which our line-level
models could barely predict. They are revised where they stand, which
our models could predict moderately well.

\textbf{Contributions.}

\begin{enumerate}
\def\labelenumi{\arabic{enumi}.}
\tightlist
\item
  Ten measurement pitfalls in studies of line lifetimes, each measured
  on one dataset and paired with a symptom by which it can be detected.
\item
  An example of pitfalls compounding: three of them together produced
  the strongest effect in an earlier version of our model.
\item
  A corrected picture of line survival: removal and revision are
  different processes, and only revision could our models rank for an
  individual line.
\item
  A checklist for authors and reviewers, and a count of how often the
  practices behind the pitfalls appear in 22 published studies.
\end{enumerate}

Section 2 places the work among related studies, §3 describes our data
and methods, and §4 presents the pitfalls. Section 5 reports what
survives the corrections, and §6 turns the pitfalls into a checklist.

\section{Related Work}\label{related-work}

\textbf{Lifetimes of code.} Survival analysis entered software
engineering at the level of projects, packages and frameworks {[}3{]},
{[}4{]}, {[}5{]}. Scanniello {[}6{]} applied Kaplan--Meier estimation
{[}7{]} to source code. Spinellis et al.~{[}1{]} moved down to lines and
tokens; they found high infant mortality in every project but no feature
that predicted longevity. Rahman and Shihab {[}2{]} studied how code
written by AI agents is modified, ending a line at any change that
\texttt{git\ blame} attributes to a later commit. Change-proneness has
also been studied for classes and files {[}8{]}, and code metrics have
been shown to lose meaning outside their project's context {[}9{]}.

\textbf{Following a line.} Any study of lines must decide when a line is
still the same line. \texttt{git\ blame} and \texttt{git\ diff} follow
text, and git's own history can mislead in other ways {[}10{]}.
Specialised tools follow other evidence: line trackers combine text and
position {[}11{]}, token-level blame follows tokens {[}12{]}, tree
differencing follows syntax {[}13{]}, refactoring detection follows
moved and renamed code {[}14{]}, and method and variable trackers follow
program elements through renames and moves {[}15{]}, {[}16{]}. The SZZ
family, which traces faults back to the lines that introduced them,
faces the same questions of identity {[}17{]}, {[}18{]}.

\textbf{Evaluating predictions.} Predicting a line's fate borrows from
defect prediction. Just-in-time models score commits by features of the
change, its files and its author {[}19{]}, and line-level variants rank
the lines of a file {[}20{]} or a commit {[}21{]}; process metrics such
as these tend to beat properties of the code {[}22{]}. This literature
has also shown how evaluation misleads. Ignoring time overstates
accuracy {[}23{]}, {[}24{]}, performance drifts over time {[}25{]},
{[}26{]}, and the validation technique changes the estimates {[}27{]}.
The data are uneven as well: large commits differ in kind from small
ones {[}28{]}, tangled changes add noise {[}29{]}, and files that change
together keep changing together {[}30{]}.

\textbf{Pitfall studies.} Kalliamvakou et al.~{[}31{]} catalogued how
GitHub data can mislead, Tantithamthavorn and Hassan {[}32{]} reported
pitfalls in defect modelling, and Kapoor and Narayanan {[}33{]} found
leakage in 294 papers across seventeen fields that adopted machine
learning. We do the same for one analysis, how long lines live, and
measure each pitfall on one dataset, before and after its correction.

\section{Data and Methods}\label{data-and-methods}

\textbf{Corpus.} We drew 120 non-forked TypeScript repositories older
than 24 months from GitHub: forty at random from the 200 most-starred in
each of three popularity strata (1,000--5,000, 5,000--20,000 and over
20,000 stars). We chose TypeScript because the author has written it
professionally for many years. That made it practical to check inferred
events by reading the code, which our validation required. Its
tree-sitter grammar is also mature. We traversed the histories with
PyDriller {[}34{]}, which visits commits from oldest to newest and gives
merge commits no changes. The histories hold 588,902 commits and 32.5
million line births. We do not record comments or lines shorter than
five characters, and when one of several identical lines in a file is
deleted, we assign the deletion to the oldest.

\textbf{Line identity.} Before we count a line as dead, we must know
that it does not live on elsewhere. Each deleted line in a commit is
therefore matched in five stages, from the most certain to the least:

\begin{enumerate}
\def\labelenumi{\arabic{enumi}.}
\tightlist
\item
  an exact match within the file;
\item
  an exact match elsewhere in the repository, which counts as a move and
  keeps the line's identity;
\item
  an assignment within the file between lines of the same AST node type
  whose similarity is at least 0.6, solved as an assignment problem
  (greedily for very large edits);
\item
  the same, greedily, across files;
\item
  a similarity of at least 0.9 within the file, regardless of node type.
\end{enumerate}

Similarity combines the overlap of the two lines' tokens (0.7 times
their Dice coefficient) with the similarity of their characters (0.3
times the Ratcliff--Obershelp ratio). A line matched in the last three
stages is \emph{rewritten}; a line not matched at all \emph{dies}. A
rewrite records the new version as a new line, so 19.9\% of all births
are rewrites of earlier lines. The survival models therefore follow
versions of lines. Sections 4.1 and 4.5 also follow \emph{lineages}: a
line as first written, through all its rewrites, to the deletion of its
last version.

\textbf{Validation.} We checked the matcher in two ways. First, the
author checked 200 lines against their repositories, days after the data
were mined. None of the 100 lines the pipeline counts dead was alive, so
false deaths are at most 3\% (one-sided 95\% bound). Of the 100 lines it
counts alive, 30 could not be found in the repository's latest version.
On closer reading, 18 of them were deaths the pipeline had missed, most
of them in files that could no longer be found; the others had been
rewritten, or sat in renamed files.

The first check has a blind spot. If the matcher wrongly pairs a deleted
line with an unrelated new line, the deleted line looks rewritten rather
than dead, and a status check cannot tell. So we also checked the pairs
themselves. The author then labelled 100 of the matcher's decisions
blind against the commit diffs, judging for each whether the two lines
are really the same line: 40 rewrites with similarity just above the
threshold (0.6--0.7), 20 rewrites above 0.7, and 40 moves. An AI
assistant reviewed the labels afterwards, and four were changed with the
author's agreement. The weak spot is the threshold. There, 13 of the 40
rewrites hide a death; above it, and among moves, none does. Only one
rewrite in eight lies near the threshold, so about 4\% of all rewrites
hide a death. Taking both checks together, the matcher finds about three
deaths in four (73\%; 95\% CI 0.64--0.80), or 63\% in the most
pessimistic count. In short, it misses some deaths but almost never
invents one.

\textbf{Fates.} We classify a line's first event by the size of the edit
to its file (Table 1). A line is \emph{removed} if it is deleted
together with its file, or in a commit that deletes more than twenty
lines of that file. It is \emph{revised locally} if it is deleted or
rewritten in a commit that deletes, or rewrites, at most twenty lines of
the file; a rewrite that only reformats does not count. Deletions and
rewrites are counted separately, and a line's file is the path where it
was born.

\begin{table*}[t]
\centering
\caption{First event of the lines of the first prediction sample; on the second, every share is within 0.2 points.}
\footnotesize
\renewcommand{\arraystretch}{1.2}
\begin{tabular}{@{}ll@{}}
\toprule
Fate & Share of lines \\
\midrule

\bottomrule

Removed with its file & 15.3\% \\
Removed in a bulk edit (\textgreater{} 20 lines of the file) & 10.3\% \\
Deleted in a local edit (1--5 / 6--20 lines) & 3.9\% / 4.3\% \\
Rewritten in a local edit, not formatting-only & 12.3\% \\
Rewritten in a bulk edit & 5.0\% \\
Reformatted only & 2.6\% \\
No event before observation ended & 46.3\% \\
\end{tabular}\end{table*}

\textbf{Survival analysis in brief.} Survival analysis models the time
until an event. Its key device is \emph{censoring}. A line still alive
when the data end, or one that left in a way we do not count, is
censored: it contributes the time during which it was observed without
the event, but no event. The Cox model {[}35{]} estimates for each
feature a \emph{hazard ratio}, the factor by which the feature
multiplies the rate of the event; below one, the feature protects.
Predictions are compared by \emph{concordance} {[}36{]}, the share of
pairs of lines whose predicted risks are in the same order as their
lifetimes. Concordance is 0.5 for a random ranking and 1 for a perfect
one. Computed within a project, it compares only lines of the same
project.

\textbf{Two analyses.} We ran two kinds of analysis on different
samples. The \emph{survival models} explain. They are fitted with
lifelines {[}37{]} and R \texttt{survival} {[}38{]} to a hash sample of
349,510 lines, with fifteen interpretable features: nesting depth, four
flags for the enclosing block, five groups of AST node types
(declarations, expressions, imports and exports, type annotations, and
other nodes, each compared with control-flow statements), the number of
tokens (log), the distance to the nearest \texttt{TODO}, how rare the
line's text is in the corpus (\texttt{lidf}, its inverse document
frequency), and whether the line was committed at night or at a weekend.
Their event is removal. Local revision is a competing event with its own
model {[}39{]}, and any other first event censors the line. The models
are not penalised, and all confidence intervals and p-values are
clustered by repository (§4.10).

The \emph{prediction models} forecast. They are gradient-boosted
survival models (XGBoost {[}40{]} with a Cox objective) and a
ridge-penalised linear Cox model, trained on two further hash samples of
about 2.03 million lines each. They use up to 196 features known when a
line is born, describing the line, the commit that added it, its file,
its directory, its author and its repository. They are evaluated forward
in time. At each of five cutoffs from July 2022 to July 2024, a model is
trained on lines born before the cutoff and tested on lines born in the
next six months. Concordance within projects is averaged over the 41 to
61 more active repositories that have at least twenty events and twenty
non-events in the test period, weighted by their number of test lines.
Unless a sample is named, prediction results hold on both samples.
(\texttt{lidf} is counted over the whole corpus, future included; the
prediction models use its value at the line's birth.)

\section{Ten Pitfalls}\label{ten-pitfalls}

The pitfalls follow the order of an analysis, from mining the history to
drawing inferences. Each is described in the same way: the practice,
what it did to our results (or would have done), why, how to notice it,
and the remedy. Table 4 at the end of the section summarises them, with
their effect on our data.

\subsection{Mining: Counting Moved and Edited Lines as Dead
(P1)}\label{mining-counting-moved-and-edited-lines-as-dead-p1}

\textbf{The practice.} A line counts as dead when it is missing from the
next revision. Take a function moved from \texttt{utils.ts} to
\texttt{format.ts}, or a variable renamed from \texttt{cnt} to
\texttt{count}. Nothing has left the program, yet a plain
\texttt{git\ diff} reports each affected line as deleted, and a new one
as added. \texttt{git\ blame} can do better: with \texttt{-M} and
\texttt{-C} it follows lines moved within a file or copied from other
files, and with \texttt{-w} it ignores changes in whitespace. But blame
answers a different question. It explains the lines present in one
revision, so a deleted line, the event whose time we need, never appears
in its output. And blame matches only unchanged lines, so a line edited
at all, even by one renamed identifier, becomes a new line.

\textbf{What it would do.} We guarded against this pitfall from the
start, so we can only estimate its cost. Our matcher (§3) finds that 8.3
million deleted lines live on: 1.8 million were moved to another file
and 6.5 million were rewritten. Counted as deaths, they would add three
false deaths for every four real ones. On the survival sample, half of
all lines as first written would end within 1.3 years. Followed through
their rewrites, half are in fact deleted within 3.0 years. The model of
removal would change as well. Token count, which protects a line (hazard
ratio 0.839), would appear to endanger it (1.124), and a model trained
on the naive events would rank true deaths worse (concordance 0.541
instead of 0.590). These figures understate the problem, because our
miner does not record lines that reappear unchanged in the same file.

\textbf{Why.} The lines rewritten soonest are the ones edited in place:
long lines that formatters rewrap, and import lists that change whenever
what they import changes. A naive diff kills exactly these lines early.
The features that mark them, length among them, then take on the risk of
revision.

\textbf{How to notice.} Moves come in bulk. Two thirds of all moves in
our corpus happened in 0.6\% of the commits, each of which moved at
least a hundred lines. Deaths that spike in such commits are the main
symptom. Another is an effect that flips sign when rewrites are taken
out of the event.

\textbf{Remedy.} Match deleted lines to added ones, including edited
ones, before counting deaths, and validate the matcher by reading diffs.
The remaining error then has a known direction. A false match hides a
death, so a validated matcher misses some deaths but invents almost none
(§3).

\subsection{Outcome: One Event for Every Way a Line Ends
(P2)}\label{outcome-one-event-for-every-way-a-line-ends-p2}

\textbf{The practice.} A line has a single way to end: it is deleted,
or, in the definition \texttt{git\ blame} suggests, it changes in any
way. Rahman and Shihab {[}2{]} use the second definition and call a
line's death ``a neutral term denoting any modification event''. It fits
their question, how long code written by agents stays as written. It
does not fit a question about removal.

\textbf{What happened.} Our earlier survival models and our first
prediction models used any deletion as the event. Forward in time, they
ranked lines within a project barely better than chance (0.54), and we
concluded that a line's future cannot be predicted from what is known at
its birth. That conclusion was wrong. Table 1 shows that lines leave in
different ways. A quarter leave with their file or in bulk, through a
decision about a module or a feature rather than about the line. A fifth
are revised in small, local edits. Table 2 gives the same features five
definitions of the event. Any change, any deletion and removal are all
ranked within projects at 0.54--0.57. Local revision is ranked at 0.71,
and at 0.76 when the edit touches at most five lines. The AUC at one
year, which does not depend on how common the event is, tells the same
story.

\begin{table*}[t]
\centering
\caption{One feature set, five definitions of a line's end: share of test lines with the event (event rate), concordance and AUC at 365 days forward in time, mean over five cutoffs, first and second sample.}
\footnotesize
\renewcommand{\arraystretch}{1.2}
\begin{tabular}{@{}
  >{\raggedright\arraybackslash}p{(\linewidth - 8\tabcolsep) * \real{0.2000}}
  >{\raggedright\arraybackslash}p{(\linewidth - 8\tabcolsep) * \real{0.2000}}
  >{\raggedright\arraybackslash}p{(\linewidth - 8\tabcolsep) * \real{0.2000}}
  >{\raggedright\arraybackslash}p{(\linewidth - 8\tabcolsep) * \real{0.2000}}
  >{\raggedright\arraybackslash}p{(\linewidth - 8\tabcolsep) * \real{0.2000}}@{}}
\toprule
The line ends when it is\ldots{}
 & Event rate
 & Overall
 & Within projects
 & AUC@365
 \\
\midrule

\bottomrule

changed in any way (\texttt{git\ blame}) & 59\% & 0.617 / 0.603 & 0.565
/ 0.546 & 0.656 / 0.651 \\
deleted in any way & 40\% & 0.605 / 0.623 & 0.540 / 0.539 & 0.649 /
0.658 \\
removed with its file or in bulk & 31\% & 0.628 / 0.669 & 0.548 / 0.559
& 0.655 / 0.701 \\
revised locally (\(\le\) 20 lines of the file) & 22\% & 0.739 / 0.738 &
\textbf{0.712 / 0.710} & 0.763 / 0.763 \\
revised locally (\(\le\) 5 lines) & 12\% & 0.786 / 0.785 & \textbf{0.760
/ 0.760} & 0.807 / 0.805 \\
\end{tabular}\end{table*}

\textbf{Why.} Removal and revision are different processes. When they
are pooled, the larger one sets the result: the pooled event is as
unpredictable as bulk removal, and the signal of local revision is lost
in it.

\textbf{How to notice.} Before fitting any model, tabulate how lines
leave: the share that leaves with its file, the size of the commits in
which lines die, and the share of rewrites that only reformat. Nothing
in our models warned us; the warning was in this table, which we had not
made.

\textbf{Remedy.} Model the ways of leaving separately, as
\emph{competing risks} {[}39{]}: each way gets its own model, and a line
that leaves in another way is censored at that moment.

\subsection{Features: Text Metrics That Measure Size
(P3)}\label{features-text-metrics-that-measure-size-p3}

\textbf{The practice.} Features computed from a line's text enter the
model as properties of the line. One of ours was the Shannon entropy of
the line's characters, meant to capture how varied or repetitive the
line is.

\textbf{What happened.} In an earlier version of our model, entropy was
the strongest protective feature: each unit of log entropy cut the
hazard of deletion by 44\% (hazard ratio 0.56), with a p-value below
\(10^{-130}\). A p-value that small is itself a warning sign: it comes
from treating hundreds of thousands of lines as independent observations
(P10). Even the word ``strongest'' was shaky, since the features were on
different scales; per standard deviation, entropy's effect (0.93) was
weaker than that of import statements or of lines inside functions. Then
a reader outside the project wrote to us that entropy was a red herring,
a stand-in for size. The reader was right.

\textbf{Why.} Entropy counts the bits needed per character. A line of
\(n\) characters can reach at most \(\log_2 n\) bits, and only if no
character repeats. \texttt{return;} has 2.5 bits out of a possible 2.8;
the 33-character line
\texttt{import\ \{\ useState\ \}\ from\ \textquotesingle{}react\textquotesingle{};}
has 3.9 out of 5.0; a 79-character line that totals the prices of a list
has 4.3 out of 6.3. Longer lines simply have room for more distinct
characters. A line's entropy is therefore the logarithm of its length
minus a term for repetition. Across our lines, the logarithm of length
alone explains 75.2\% of the variance of entropy. The rest, repetition,
has no effect at all: with length in the model, the residual of entropy,
what remains of it after regression on log length, has a hazard ratio of
1.009 per standard deviation. What looked like an effect of varied text
was an effect of long text, which is how size confounds code metrics in
general {[}41{]}. Length, in turn, mixes two things. A longer line holds
more tokens, more content, and that protects it (0.90 per standard
deviation of token count). But long lines are also rewrapped by
formatters more often, and the censoring error of P5 credited every
rewritten line with years of survival. With that error corrected and
standard errors clustered by repository, neither entropy nor character
length is significant; token count remains.

\textbf{How to notice.} A feature derived from the text is stronger than
the plain measurement it is derived from.

\textbf{Remedy.} Put size into the model explicitly, test what remains
of the derived feature once size is there, and measure size by what the
code contains: tokens rather than characters.

\subsection{Time I: Ignoring Lines Still Alive
(P4)}\label{time-i-ignoring-lines-still-alive-p4}

\textbf{The practice.} Which features make lines live longer? A simple
approach takes the lines that have already died and regresses their
lifetimes on their features by ordinary least squares. Another
classifies lines by whether they died within a fixed time, say a year.
Both share one flaw: they cannot use a line that is still alive when the
data end. In our survival sample, that is three lines in four. Spinellis
et al.~{[}1{]} described lifetimes with Kaplan--Meier and Weibull
models. To look for features that predict longevity, however, they
regressed observed lifetimes in each project. They did not say how lines
still alive were handled, and they found no such features.

\textbf{What it would do.} We fitted both regressions to the same lines
as a Cox model of removal (Table 3). Regressed on the lines that were
removed, the fifteen features explain 3.0\% of the variance of log
lifetime, and only one coefficient is significant: ``no feature predicts
longevity''. The Cox model, which uses all lines, finds six effects.
Three of the six point the other way in the regression, though not
significantly: expression statements, which the Cox model finds removed
faster, appear to live 77\% longer. Counting censored lines as if they
had been removed ranks lines worst of all. A classifier with a one-year
horizon would discard the 41\% of lines observed for less than a year.

\begin{table*}[t]
\centering
\caption{Observed-lifetime regression against a survival model of removal on the same lines of the survival sample, fitted on a random 70\% of lines and scored on 30\% (a descriptive comparison); significance with standard errors clustered by repository. ``Opposite to Cox'' counts point estimates that point the other way: none of the three is significant for removed lines only, and one of the two (token count) when censored lines are counted as removed.}
\footnotesize
\renewcommand{\arraystretch}{1.2}
\begin{tabular}{@{}
  >{\raggedright\arraybackslash}p{(\linewidth - 10\tabcolsep) * \real{0.1667}}
  >{\raggedright\arraybackslash}p{(\linewidth - 10\tabcolsep) * \real{0.1667}}
  >{\raggedright\arraybackslash}p{(\linewidth - 10\tabcolsep) * \real{0.1667}}
  >{\raggedright\arraybackslash}p{(\linewidth - 10\tabcolsep) * \real{0.1667}}
  >{\raggedright\arraybackslash}p{(\linewidth - 10\tabcolsep) * \real{0.1667}}
  >{\raggedright\arraybackslash}p{(\linewidth - 10\tabcolsep) * \real{0.1667}}@{}}
\toprule
Model
 & Lines used
 & R²
 & Concordance
 & Significant effects
 & Opposite to Cox
 \\
\midrule

\bottomrule

Cox proportional hazards & all & --- & 0.612 & 6 of 15 & --- \\
OLS of log lifetime, removed lines only & 26\% & 0.030 & 0.596 & 1 of 15
& 3 of 6 \\
OLS, censored lines counted as removed & all & 0.014 & 0.569 & 4 of 15 &
2 of 6 \\
\end{tabular}\end{table*}

\textbf{Why.} Looking only at lines that died selects them by their
outcome and by how long they could be watched. A line born late in a
repository's history can appear among the dead only if it died quickly.
Any feature linked to when lines are born, or to how fast they die, then
shifts the observed lifetimes for reasons unrelated to its effect.

\textbf{How to notice.} A small R², and effects whose directions
disagree with a survival model or change when censored lines are added.

\textbf{Remedy.} Use a survival model. If a fixed-horizon classifier is
unavoidable, report the share of lines it discards, and check that this
share does not depend on the features used to rank risk.

\subsection{Time II: Censoring That Picks the Question
(P5)}\label{time-ii-censoring-that-picks-the-question-p5}

\textbf{The practice.} A line can leave in more than one way, and each
way is a competing risk. Take a line rewritten in its second year, in a
repository observed for ten. The original line no longer exists and can
no longer be deleted, but the analysis must still decide what to do with
it. Censored at the rewrite, it feeds a \emph{cause-specific} model,
which asks what makes lines be removed. Kept at risk until the
repository's last commit, eight years later, it feeds a
\emph{Fine--Gray} model {[}42{]}, which asks how many lines are removed
in the end. Such a line can never be removed, so it counts as a line
that survived; the model then describes the share of all lines that end
up removed, which is smaller wherever rewriting comes first, rather than
how fast the lines that still exist are removed. The second option is
also the simpler query, since it censors every line not yet deleted at
the last commit.

\textbf{What happened.} Our paper asked the first question, but our
loaders answered the second: the join to the table of rewrites was
missing. We found this only by reading the SQL against the definition
stated in the paper. A fifth of the sample (19.8\%) was affected, and
each of these lines stayed at risk for 1,081 days on average after it
could no longer be deleted. Epidemiologists call such follow-up
\emph{immortal time}; counted as survival, it produces spurious
protective effects {[}43{]}. Fig. 2 shows what it did to us. With the
error, the protective effect of token count grew stronger with age (from
0.755 in the first ninety days to 0.691 after a year). Corrected, it
fades (from 0.789 to 0.896, no longer significant). Import and export
statements older than a year looked well protected (0.499); corrected,
the effect is not significant. An explicit Fine--Gray model reproduces
the erroneous estimates: in effect, the loaders had fitted a Fine--Gray
model.

\begin{figure}
\centering
\pandocbounded{\includegraphics[keepaspectratio,alt={Hazard ratios of removal by age interval when a rewritten line is censored at the repository's last commit (dashed) and at its rewrite (solid, with a 95\% confidence band clustered by repository).}]{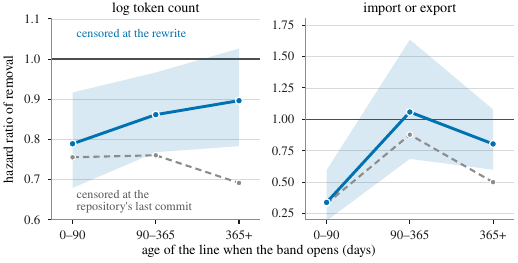}}
\caption{Hazard ratios of removal by age interval when a rewritten line
is censored at the repository's last commit (dashed) and at its rewrite
(solid, with a 95\% confidence band clustered by repository).}
\end{figure}

The same rule produced the most quoted number of an earlier version of
our analysis: that more than half of all lines are never deleted. As a
statement about lines as written, which a rewrite ends, it was fair: on
this risk set, one minus the Kaplan--Meier estimate is in effect the
cumulative incidence of deletion (48.9\%, against 49.1\% by the
Aalen--Johansen estimator). We read it, however, as a statement about
lines, and half of all lines, followed through their rewrites, are
deleted within 3.0 years. The Kaplan--Meier median we first put in its
place, censoring rewrites, was no better: it describes a world in which
no line is ever rewritten.

\textbf{Why.} On the Fine--Gray risk set, a rewritten line still counts
as at risk but can no longer be removed. A feature that makes rewriting
likely, such as length (P3), therefore lowers the hazard, and lowers it
more as rewritten lines pile up with age. The model was doing what it is
meant to do. Our mistake was to read a statement about incidence as a
property of lines, and an effect that grows with age invites exactly
that reading.

\textbf{How to notice.} A protective effect that grows steadily with
age, for a feature that makes a competing event more likely.

\textbf{Remedy.} Decide whether the question is about causes or about
how many lines are affected, and censor accordingly. Then read the
loaders against the stated definition; that is how we found our error.

\subsection{Model I: One Hazard Ratio for a Whole Life
(P6)}\label{model-i-one-hazard-ratio-for-a-whole-life-p6}

\textbf{The practice.} The Cox model reports one hazard ratio per
feature for a line's whole life. This assumes \emph{proportional
hazards}: a feature multiplies the risk by the same factor at every age.

\textbf{What it would do.} We tested the assumption from the start, so
we can only show what a single number would have said. On all ages,
declarations appear to be removed 17\% more slowly than control-flow
lines (0.831, not significant), and expression statements 26\% faster
(1.262).

\textbf{What the ages show.} The Schoenfeld test {[}44{]} rejects
proportional hazards for fourteen of the fifteen features. Landmark
models, fitted separately to lines still at risk at 0--90 days, 90--365
days and after a year, show how the effects change. A declaration is
removed at half the rate of a control-flow line in its first ninety days
(0.487), and no differently afterwards. The single 0.831 describes no
age at all. Expression statements run the other way: no different at
first, then removed faster (1.817 between ninety days and a year).

\textbf{Why.} Part of the change with age is not about lines at all, and
here P6 meets P7. Lines of fast-removing projects leave first, so the
mix of projects among surviving lines changes with age, and a pooled
model sees this as effects that change with age. When each repository
gets its own baseline, the test still rejects ten features of fifteen,
but the evidence shifts: for conditionals and night commits it falls
more than tenfold, while for weekends and lines inside functions it
grows. Some of the change is projects mixed together, and some is real.
(These tests use model-based variances, which P10 shows to be too small,
so we rely on the landmark estimates and their clustered intervals.)

\textbf{How to notice.} The Schoenfeld test rejects most features. On
hundreds of thousands of lines it also flags departures of little
practical weight, so read a rejection as ``the effect changes'', not as
``the model is useless''.

\textbf{Remedy.} Estimate effects by age, with landmark or time-varying
models, with and without the project in the model. Check the censoring
first: age-specific estimates are exactly where immortal time (P5) shows
up as a trend.

\subsection{Model II: Treating All Projects as One
(P7)}\label{model-ii-treating-all-projects-as-one-p7}

\textbf{The practice.} All lines are treated as one population, and the
features are expected to absorb the differences between projects.

\textbf{What nearly happened.} In a pooled model with naive standard
errors, lines inside loops appeared protected (0.927, p = 0.002) and
lines in \texttt{try}/\texttt{catch} blocks endangered (1.069, p =
0.02). We nearly reported both. But projects differ widely. A
\emph{shared frailty} model {[}45{]}, {[}38{]}, a Cox model with a
random effect for each project, gives each project its own baseline
hazard, drawn from a common distribution. In ours, the baseline hazards
of the projects at the 10th and 90th percentiles are 12.5 times apart.
With the frailty in the model, both move to one, which means no effect,
or past it (1.030 and 0.937). Standard errors clustered by repository
(P10) would not have found them significant in the first place.
Prediction is affected too. At one of our cutoffs, a single repository
held over a third of the test deaths (37.5\%); overall concordance of
any deletion there was 0.68--0.69, and 0.59--0.61 without that
repository.

\textbf{Why.} Differences between projects that are unrelated to the
features would only pull pooled hazard ratios towards one {[}46{]}. Ours
moved away from one: eleven of the fifteen were further from one in the
pooled model than with the frailty. The features are correlated with the
project, so the pooled model credits them with the project's rate of
removal. Projects that use loops or error handling heavily differ in how
much code they remove, and the pooled model attributed that difference
to the syntax. Overall concordance, likewise, rewards knowing which
project is removing code.

\textbf{How to notice.} Effects shrink or vanish once a project frailty
or stratification is added. Overall concordance is far above concordance
within projects.

\textbf{Remedy.} Fit a frailty or stratify by project, report prediction
within projects, and state how much of the test set the largest project
holds.

\subsection{Evaluation I: Splits That Leak the Outcome
(P8)}\label{evaluation-i-splits-that-leak-the-outcome-p8}

\textbf{The practice.} When the unit of analysis is small and the data
cover one period, lines are split into training and test sets at random
or by file. Every feature is computed from what was known at the line's
birth, so nothing from the future seems able to reach the model.

\textbf{What happened.} A diagnostic run of our first prediction model,
a classifier of death within ninety days on lines of one half-year split
by file, reached an AUC of 0.92. Nothing in our survival models could
explain so sharp a result, and we did not believe it. Fig. 1 shows what
was going on. It keeps the model and the lines fixed and changes only
how the lines are separated. For removal, concordance is 0.83 when lines
are split by file, about 0.75 when they are split by the commit that
added them or by directory, and 0.72 when the model is evaluated forward
in time on the same half-year. The leak inflates concordance by about
0.11. Local revision loses only about 0.06. Evaluating within projects,
the remedy for P7, does not stop the leak: under the file split, removal
is ranked within projects at 0.71, against 0.61 forward in time. A split
by repository gives the lowest numbers of all, but it answers another
question, how a model transfers to projects it has never seen.

\textbf{Why.} The usual explanation of such leaks is that the model sees
the future {[}23{]}, {[}24{]}. Here every feature is known at birth;
what the two halves share is the outcome. In the terms of Kapoor and
Narayanan {[}33{]}, training and test data are not independent. Lines
leave together: of the lines born in one test half-year that were
deleted within a year, 44\% went in its 100 largest events, each a
repository and a moment of death. Shared deaths are not the main
channel, though, although an earlier draft of this section said they
were. Keeping all lines of one event on the same side lowers removal
only to 0.77. The commit split, which still shares events, comes close
to forward evaluation (0.75 against 0.72). What a file split adds is the
shared birth. Lines added by one commit share their context and much of
their fate, and a file split puts them on both sides.

\textbf{How to notice.} Accuracy falls steeply as the split coarsens
from file to commit, and further under forward evaluation. A second
warning is instability forward in time: removal concordance ranges from
0.55 to 0.72 across our five cutoffs, depending on which projects happen
to be removing code, while revision stays between 0.70 and 0.76.

\textbf{Remedy.} Evaluate forward in time, within projects, over several
cutoffs, and report the spread.

\subsection{Evaluation II: Judging a Ranking by Its Top
(P9)}\label{evaluation-ii-judging-a-ranking-by-its-top-p9}

\textbf{The practice.} A ranking is judged by its top: of the lines a
model rates riskiest, how many had the event.

\textbf{What happened.} A draft of this analysis reported that of the
forty test lines with the highest predicted risk of local revision, 80\%
were revised, against 15\% of forty random lines. Then we read the forty
lines. Thirty-seven were leftover debugging output, lines too long for
the formatter, long import and export lists, or version constants, and
they came from a ranking of all projects at once. No developer needs a
model to learn that a stray debugging statement will go or that an
overlong line will be wrapped. With such obvious cases set aside (15\%
of lines), and lines ranked within each project, the top 1\% of a
project's new lines is revised within a year at 49\%, against a base
rate of 21\%: 2.3 times as often. The best ranking from version history
alone, by how few lines the birth commit added to the file, reaches 1.7
times the base rate. Of forty random lines from the non-obvious top 1\%,
22 were revised within a year, mostly in small follow-ups: a label
renamed together with its key, a type widened.

\textbf{Why.} A count of hits at the top rewards the easiest hits, and
the easiest hits are edits no one needs predicted. Ranking all projects
together makes it worse: ranked within each project, the top 5\% is
already half ordinary lines (55\%).

\textbf{How to notice.} Read the top of the list.

\textbf{Remedy.} Rank within projects, set aside what a linter or
formatter already flags, and compare with the simplest ranking from
version history.

\subsection{Inference: Treating Lines as Independent
(P10)}\label{inference-treating-lines-as-independent-p10}

\textbf{The practice.} Each line is one observation, and a model fitted
to hundreds of thousands of lines reports standard errors to match.

\textbf{What happened.} Every survival model in earlier versions of our
analysis did this. We checked once, clustering lines by the commit that
added them. The intervals widened three- to tenfold, and we recorded the
check without acting on it. Clustered by repository with a sandwich
estimator {[}47{]}, which lets lines of one repository be correlated,
the confidence intervals of the fifteen effects on removal are 2.4 to 19
times as wide as the naive ones. Only six effects remain significant; a
jackknife that leaves out one repository at a time keeps five, and a
Holm correction for fifteen tests keeps four. With both corrections at
once, only token count remains. Night-time commits, with a naive p-value
below \(10^{-100}\), are not significant (p = 0.22). The correction
reaches the other pitfalls too: the significant reversals of the
lifetime regression (P4) disappear, the naive diff (P1) keeps one, and
entropy (P3) is not significant with or without the censoring error.

\textbf{Why.} Lines are born together, die together (P8) and share their
project's rate of removal (P7). For an effect that varies between
projects, the effective sample size is closer to the number of projects
than to the number of lines. With little more than a hundred projects
and removals that arrive in bulk, a naive standard error is too small by
an order of magnitude.

\textbf{How to notice.} Clustered intervals are many times wider than
naive ones.

\textbf{Remedy.} Cluster standard errors by project, or by the coarsest
unit whose lines share events. For a frailty model, read the point
estimates: its model-based standard errors still treat the lines of a
project as independent once the frailty is known.

\begin{table*}[t]
\centering
\caption{The pitfalls, their effect on our data, their symptoms and their remedies.}
\footnotesize
\renewcommand{\arraystretch}{1.2}
\begin{tabular}{@{}
  >{\raggedright\arraybackslash}p{(\linewidth - 6\tabcolsep) * \real{0.2500}}
  >{\raggedright\arraybackslash}p{(\linewidth - 6\tabcolsep) * \real{0.2500}}
  >{\raggedright\arraybackslash}p{(\linewidth - 6\tabcolsep) * \real{0.2500}}
  >{\raggedright\arraybackslash}p{(\linewidth - 6\tabcolsep) * \real{0.2500}}@{}}
\toprule
Pitfall
 & On our data
 & Symptom
 & Remedy
 \\
\midrule

\bottomrule

P1 Moved lines as dead & median life 1.3 years instead of 3.0; token
count flips from protective to risky & deaths spike in commits that move
code; effects flip when rewrites leave the event & match moves and
rewrites; validate on read diffs \\
P2 One event for all endings & within-project concordance 0.56 instead
of 0.71 & most ``deaths'' leave with their file or in bulk & tabulate
fates; model removal and revision as competing risks \\
P3 Text metrics measuring size & the strongest feature measured length &
a text feature beats plain length & add size explicitly; test the
residual; count tokens \\
P4 Ignoring the living & 1 significant effect instead of 6; three
directions reversed & small R²; directions disagree with a survival
model & survival model; report what a horizon discards \\
P5 Censoring picks the question & protection that fades with age appears
to grow & protective effect growing with age, for features linked to a
competing event & censor at the competing event for causes, Fine--Gray
for incidence; check loaders against text \\
P6 One ratio for a whole life & one hazard ratio that describes no age &
Schoenfeld test rejects most features & landmark or time-varying
effects, with and without the project; censoring first \\
P7 All projects as one & two syntax effects that were project effects &
effects shrink under a project frailty; overall concordance far above
within-project & frailty or stratification; within-project metrics;
share of the largest project \\
P8 Leaky splits & concordance 0.83 by file, 0.72 forward in time (within
projects 0.71 and 0.61) & accuracy falls as splits coarsen, even within
projects; unstable across cutoffs & forward in time, within projects,
over several cutoffs \\
P9 Judging by the top & 37 of the top 40 lines obvious & the top of the
list is obvious & rank within projects; set aside linter targets;
compare with history \\
P10 Lines as independent & 9 of 15 significant effects vanish &
clustered intervals many times wider than naive ones & cluster by
project; count projects, not lines \\
\end{tabular}\end{table*}

\section{What Survives the
Corrections}\label{what-survives-the-corrections}

With the pitfalls corrected, two findings remain, one for each way a
line leaves.

\textbf{Removal belongs to modules and projects more than to lines.}
Most deaths are mass deaths. Of 11.0 million, 45\% remove a whole file,
and a further 19\% happen in commits that delete more than a hundred
lines of the file. Of the lines born in one commit that later die, two
thirds (67.6\%) die together in one commit. Removal does depend on a
line's age (P6) and above all on its project (P7). Within a project,
however, our models could barely rank it: concordance stays between 0.54
and 0.57 under every definition that includes removal, on both samples,
and its bootstrap interval includes chance at three of five cutoffs.
Missed deaths do not explain this. Hiding a third of the removals of
whole files, where most missed deaths are, moves within-project
concordance by at most 0.004. Nor did other models do better when we
tried them on any deletion: a linear Cox model and embeddings of the
line's text from a code model {[}48{]} added nothing.

\textbf{Local revision can be ranked, moderately.} Of the lines observed
for a full year, 28.6\% are revised within it. The same features rank
local revision within projects at 0.71 on both samples, and a linear Cox
model does as well. The result is not an artefact of the definition. In
a small file almost any edit is local, so file size alone ranks revision
at 0.63; but within strata of project and file size, where size cannot
help, the model keeps 0.66, against 0.61 for version history alone. It
holds on lines as first written, without the rewrites (0.72), and
without long lines, imports and debugging output. Most of the signal
lies in how and where a line was born. The commit that added it, and
especially how few lines that commit added to the file, carries a
quarter of the model's attributions, followed by the file, the
repository, the line itself and the author. Patterns that developers
read as bad practice point both ways: casts to \texttt{any} and
suppressed compiler errors are revised more often, \texttt{var}
declarations and hard-coded URLs less often. Once the birth context is
known, they add almost nothing (0.003), in line with the finding that
process measures beat properties of the code {[}22{]}.

\textbf{For tools,} three consequences follow. Predict revision rather
than removal. Calibrate to the project and to the line's age. Evaluate
forward in time, within projects. What such a tool would show is modest:
freshly patched spots in ordinary code that the next edits are likely to
touch. Whether developers would find that worth acting on, we have not
tested.

\section{A Checklist}\label{a-checklist}

Table 4 turns the pitfalls into checks that an author can run and a
reviewer can ask for. Whether such a checklist is needed depends on how
common the practices behind the pitfalls are, so we looked for them in
published studies.

\textbf{How common the practices are.} We searched OpenAlex for studies
of how long code elements last: lines and tokens, clones, smells,
self-admitted technical debt, static-analysis issues, logging
statements, feature toggles, dead methods and code written by AI agents.
We also followed the citations of ten of them. Of 1,542 records, 50
studies qualified on their abstracts. We could read only those with an
open copy, such as a preprint or an author's version: 25 of the 50, of
which 22 measure lifetimes. Studies available only from their publishers
are therefore under-represented. We coded those 22 against Table 4 with
an AI assistant; the artifact lists all 50, with the coding and the
passages it rests on.

Survival methods are in wide use. Fifteen of the 22 studies used them,
and only three, all among the earliest, computed how long clones live
from those that had disappeared {[}49{]} (P4). All four Cox models were
checked for proportional hazards, although one study kept its model as
an average effect after every covariate failed the test {[}2{]} (P6).
The weak points lie before and after the model. Nine studies say how
they follow elements through renames and moves, three do so in part, one
relies on \texttt{git\ blame}, and seven do not say (P1). Twenty of the
22 let one event stand for every way an element ends (P2). A debt
comment, for example, counts as removed when its file is deleted
{[}50{]}, although such removals are often accidental {[}51{]}; only one
study kept a dead method's removal apart from its revival {[}52{]}.
Eight studies pooled the elements of many projects without a project
term (P7), and ten compared groups with tests that treat elements as
independent {[}53{]} (P10); none clustered. None of the three predictive
models was evaluated forward in time (P8). Several of these choices fit
the question their authors asked (P2), and the checklist asks only that
they be made knowingly.

\section{Threats to Validity}\label{threats-to-validity}

\textbf{Construct.} The boundary between local and bulk edits (twenty
lines) and the rule for formatting-only rewrites are choices; with a
boundary of five lines, the results are stronger. Local revision also
includes follow-up edits of the same piece of work: 11.5\% come within a
day of the line's birth. The matcher was checked by one annotator, and
the AI review of the match labels was not blind (§3). False matches hide
2--7\% of recorded deaths (95\% upper bound 16\%), so deaths are
undercounted rather than overcounted. Missed deaths barely move the
rankings (§5), but we did not test how they bias hazard ratios, in
particular for rare groups of lines. Another 1\% of rewrites get the
wrong successor, which misdirects lineages but hides no death. In three
small repositories, changing the matcher's thresholds changes death
counts by 6--17\%. Finally, histories are traversed in date order and
merge commits contribute no changes, so lines edited differently on
parallel branches may produce false deaths and births; this may explain
some missed deaths in files that could no longer be found.

\textbf{Internal.} The ``before'' numbers of P3 come from our earlier
pipeline, while those of P5 rerun the earlier censoring rule inside the
corrected analysis, so the size of each pitfall depends on which other
corrections are in place. Standard errors are clustered by repository,
the coarsest unit available. The clusters are few and unequal (the
largest holds 12.5\% of the lines), but a leave-one-repository-out
jackknife agrees on five of the six effects. The frailty model's
intervals are model-based, so for it we rely on point estimates.

Our two prediction protocols set success thresholds before the models
they judge were trained, but they were not registered with a timestamp,
and version control cannot confirm the order. The protocol for death
named the AUC at one year, at one cutoff, as its main measure. The
result, 0.650, was ``weak'' by its own thresholds, and the
within-project concordance over five cutoffs that we report instead was
chosen after that result. The protocol for local revision was written
after that result was known. Its main measure, within-project
concordance, came out at 0.712, just above its threshold of 0.70 for
``strong'', while the bad-practice patterns it expected to matter added
0.003, against the 0.02 it expected. The top-of-list analysis of P9 was
added later and is covered by neither protocol.

Competing outcomes are treated as censoring. We model cause-specific
hazards because our questions are about causes (P5). Such a model
describes removal among lines that can also be rewritten. If the
tendencies to be rewritten and to be removed share unmeasured causes, it
does not tell what removal would be if rewriting were impossible.

\textbf{External.} All repositories are TypeScript and were sampled by
popularity. The size of each pitfall depends on the corpus, but its
direction follows from its mechanism. P4 to P8 and P10 are properties of
the statistical machinery, so they should transfer to other corpora. We
checked four of them on a second corpus with no repository in common
with the first: 167 TypeScript projects that adopted spec-driven
development tools (29.9 million births; a sample of 320,956 lines, with
fourteen features, as \texttt{lidf} was not computed). With clustered
standard errors, 5 of 14 effects on removal stay significant, against 13
with naive ones (P10). The censoring error again makes the protection of
token count grow with age (P5). A naive diff shortens the median time to
deletion from 1,018 days to 430 (P1). A regression of observed lifetimes
again ranks lines worse than the survival model (0.514 against 0.556),
although there it finds as many significant effects, one of them
reversed (P4). How much of a corpus is bulk removal rather than local
revision, and how much of a naive top-of-list is formatting or imports,
are conventions of the ecosystem and may differ elsewhere (P2, P9). The
repository split suggests that the revision model transfers to unseen
projects (0.71--0.74), but we did not test this forward in time. Commits
by AI agents that leave no trace in their metadata are not
distinguished.

\section{Conclusion}\label{conclusion}

Earlier versions of this analysis said that more than half of all lines
are never deleted, and that the entropy of a line's characters is its
strongest protective feature. We concluded that a line's future cannot
be predicted from what is known at its birth. With naive standard
errors, night-time commits appeared to speed removal, at a p-value below
\(10^{-100}\), and a split by file offered an AUC of 0.92. None of this
survived. The first statement holds only for lines as written, before
any rewrite (P5). The entropy measured length (P3). The unpredictability
came from pooling two outcomes, one of which can be predicted (P2). The
night-time effect is not significant once lines are clustered by project
(P10). And the AUC came largely from lines of one commit sitting on both
sides of the split (P8).

None of these errors announced itself, but each left a symptom that can
be checked (Table 4). Once we checked for them, our TypeScript data told
a simple story. Lines are removed with the modules and projects around
them, which our line-level models could barely predict. They are revised
where they stand, which our models could predict.

\section{Data Availability}\label{data-availability}

Scripts, prediction protocols with their success thresholds, and the
numerical outputs behind every table and figure, including the
``before'' versions, are archived at \url{https://osf.io/fq7nb/} (code
under MIT, data under CC~BY~4.0). The full dataset of 32.5 million line
births (about 5 GB) is available from the author on request.

\section*{Acknowledgements}\label{acknowledgements}
\addcontentsline{toc}{section}{Acknowledgements}

The author thanks Derek M. Jones, who pointed out that character entropy
was a red herring and drew attention to the work of Spinellis et al.

Generative AI was used throughout this work. Claude (Anthropic; Claude
Opus 5 and 5.5) {[}54{]}, directed and checked by the author, wrote most
of the analysis code and ran the analyses, drafted and revised the text
of every section, checked the references against Crossref, screened and
coded the studies of §6, critiqued drafts as a simulated reviewer, and
reviewed the blind labels of the matcher sample after they were made
(§3); parts of the line miner were written with Claude and with Gemini
(Google). The status labels and the blind labels of the matches (§3)
were made by the author without AI assistance. The author takes full
responsibility for the content.

\section{References}\label{references}

{[}1{]} D. Spinellis, P. Louridas, and M. Kechagia, ``Software
evolution: the lifetime of fine-grained elements,'' \emph{PeerJ Computer
Science}, vol.~7, p.~e372, 2021.

{[}2{]} M. Rahman and E. Shihab, ``Will it survive? Deciphering the fate
of AI-generated code in open source,'' \emph{arXiv preprint
arXiv:2601.16809}, 2026.

{[}3{]} E. N. Samoladas, L. Angelis, and I. Stamelos, ``Survival
analysis on the duration of open source projects,'' \emph{Information
and Software Technology}, vol.~52, no. 9, pp.~902--922, 2010.

{[}4{]} M. Claes, T. Mens, R. Di Cosmo, and J. Vouillon, ``A historical
analysis of Debian package incompatibilities,'' in \emph{Proceedings of
the 12th Working Conference on Mining Software Repositories}, 2015,
pp.~212--223.

{[}5{]} M. Goeminne and T. Mens, ``Towards a survival analysis of
database framework usage in Java projects,'' in \emph{Proceedings of the
31st IEEE International Conference on Software Maintenance and
Evolution}, 2015, pp.~551--555.

{[}6{]} G. Scanniello, ``Source code survival with the Kaplan Meier,''
in \emph{Proceedings of the 27th IEEE International Conference on
Software Maintenance}, 2011, pp.~524--527.

{[}7{]} E. L. Kaplan and P. Meier, ``Nonparametric estimation from
incomplete observations,'' \emph{Journal of the American Statistical
Association}, vol.~53, no. 282, pp.~457--481, 1958.

{[}8{]} F. Khomh, M. Di Penta, Y.-G. Guéhéneuc, and G. Antoniol, ``An
exploratory study of the impact of antipatterns on class change- and
fault-proneness,'' \emph{Empirical Software Engineering}, vol.~17, no.
3, pp.~243--275, 2012.

{[}9{]} J. Y. Gil and G. Lalouche, ``When do software complexity metrics
mean nothing? --- When examined out of context,'' \emph{Journal of
Object Technology}, vol.~15, no. 1, pp.~2:1--25, 2016.

{[}10{]} C. Bird, P. C. Rigby, E. T. Barr, D. J. Hamilton, D. M. German,
and P. Devanbu, ``The promises and perils of mining git,'' in
\emph{Proceedings of the 6th IEEE International Working Conference on
Mining Software Repositories}, 2009, pp.~1--10.

{[}11{]} M. Asaduzzaman, C. K. Roy, K. A. Schneider, and M. Di Penta,
``LHDiff: A language-independent hybrid approach for tracking source
code lines,'' in \emph{Proceedings of the IEEE International Conference
on Software Maintenance}, 2013, pp.~230--239.

{[}12{]} D. M. German, B. Adams, and K. Stewart, ``cregit: Token-level
blame information in git version control repositories,'' \emph{Empirical
Software Engineering}, vol.~24, no. 4, pp.~2725--2763, 2019.

{[}13{]} J.-R. Falleri, F. Morandat, X. Blanc, M. Martinez, and M.
Monperrus, ``Fine-grained and accurate source code differencing,'' in
\emph{Proceedings of the 29th ACM/IEEE International Conference on
Automated Software Engineering}, 2014, pp.~313--324.

{[}14{]} D. Silva and M. T. Valente, ``RefDiff: Detecting refactorings
in version histories,'' in \emph{Proceedings of the 14th International
Conference on Mining Software Repositories}, 2017, pp.~269--279.

{[}15{]} F. Grund, S. A. Chowdhury, N. C. Bradley, B. Hall, and R.
Holmes, ``CodeShovel: Constructing method-level source code histories,''
in \emph{Proceedings of the 43rd IEEE/ACM International Conference on
Software Engineering}, 2021, pp.~1510--1522.

{[}16{]} M. Jodavi and N. Tsantalis, ``Accurate method and variable
tracking in commit history,'' in \emph{Proceedings of the 30th ACM Joint
European Software Engineering Conference and Symposium on the
Foundations of Software Engineering}, 2022, pp.~183--195.

{[}17{]} J. Śliwerski, T. Zimmermann, and A. Zeller, ``When do changes
induce fixes?'' in \emph{Proceedings of the 2005 International Workshop
on Mining Software Repositories}, 2005, pp.~1--5.

{[}18{]} G. Rosa, L. Pascarella, S. Scalabrino, R. Tufano, G. Bavota, M.
Lanza, and R. Oliveto, ``Evaluating SZZ implementations through a
developer-informed oracle,'' in \emph{Proceedings of the 43rd IEEE/ACM
International Conference on Software Engineering}, 2021, pp.~436--447.

{[}19{]} Y. Kamei, E. Shihab, B. Adams, A. E. Hassan, A. Mockus, A.
Sinha, and N. Ubayashi, ``A large-scale empirical study of just-in-time
quality assurance,'' \emph{IEEE Transactions on Software Engineering},
vol.~39, no. 6, pp.~757--773, 2013.

{[}20{]} S. Wattanakriengkrai, P. Thongtanunam, C. Tantithamthavorn, H.
Hata, and K. Matsumoto, ``Predicting defective lines using a
model-agnostic technique,'' \emph{IEEE Transactions on Software
Engineering}, vol.~48, no. 5, pp.~1480--1496, 2022.

{[}21{]} C. Pornprasit and C. Tantithamthavorn, ``JITLine: A simpler,
better, faster, finer-grained just-in-time defect prediction,'' in
\emph{Proceedings of the 18th International Conference on Mining
Software Repositories}, 2021, pp.~369--379.

{[}22{]} F. Rahman and P. Devanbu, ``How, and why, process metrics are
better,'' in \emph{Proceedings of the 35th International Conference on
Software Engineering}, 2013, pp.~432--441.

{[}23{]} M. Tan, L. Tan, S. Dara, and C. Mayeux, ``Online defect
prediction for imbalanced data,'' in \emph{Proceedings of the 37th
International Conference on Software Engineering}, vol.~2, 2015,
pp.~99--108.

{[}24{]} M. Jimenez, R. Rwemalika, M. Papadakis, F. Sarro, Y. Le Traon,
and M. Harman, ``The importance of accounting for real-world labelling
when predicting software vulnerabilities,'' in \emph{Proceedings of the
27th ACM Joint Meeting on European Software Engineering Conference and
Symposium on the Foundations of Software Engineering (ESEC/FSE)}, 2019,
pp.~695--705.

{[}25{]} S. McIntosh and Y. Kamei, ``Are fix-inducing changes a moving
target? A longitudinal case study of just-in-time defect prediction,''
\emph{IEEE Transactions on Software Engineering}, vol.~44, no. 5,
pp.~412--428, 2018.

{[}26{]} A. A. Bangash, H. Sahar, A. Hindle, and K. Ali, ``On the
time-based conclusion stability of cross-project defect prediction
models,'' \emph{Empirical Software Engineering}, vol.~25, no. 6,
pp.~5047--5083, 2020.

{[}27{]} C. Tantithamthavorn, S. McIntosh, A. E. Hassan, and K.
Matsumoto, ``An empirical comparison of model validation techniques for
defect prediction models,'' \emph{IEEE Transactions on Software
Engineering}, vol.~43, no. 1, pp.~1--18, 2017.

{[}28{]} A. Hindle, D. M. German, and R. C. Holt, ``What do large
commits tell us? A taxonomical study of large commits,'' in
\emph{Proceedings of the 2008 International Working Conference on Mining
Software Repositories}, 2008, pp.~99--108.

{[}29{]} K. Herzig and A. Zeller, ``The impact of tangled code
changes,'' in \emph{Proceedings of the 10th Working Conference on Mining
Software Repositories}, 2013, pp.~121--130.

{[}30{]} T. Zimmermann, A. Zeller, P. Weissgerber, and S. Diehl,
``Mining version histories to guide software changes,'' \emph{IEEE
Transactions on Software Engineering}, vol.~31, no. 6, pp.~429--445,
2005.

{[}31{]} E. Kalliamvakou, G. Gousios, K. Blincoe, L. Singer, D. M.
German, and D. Damian, ``The promises and perils of mining GitHub,'' in
\emph{Proceedings of the 11th Working Conference on Mining Software
Repositories}, 2014, pp.~92--101.

{[}32{]} C. Tantithamthavorn and A. E. Hassan, ``An experience report on
defect modelling in practice: Pitfalls and challenges,'' in
\emph{Proceedings of the 40th International Conference on Software
Engineering: Software Engineering in Practice}, 2018, pp.~286--295.

{[}33{]} S. Kapoor and A. Narayanan, ``Leakage and the reproducibility
crisis in machine-learning-based science,'' \emph{Patterns}, vol.~4, no.
9, p.~100804, 2023.

{[}34{]} D. Spadini, M. Aniche, and A. Bacchelli, ``PyDriller: Python
framework for mining software repositories,'' in \emph{Proceedings of
the 26th ACM Joint European Software Engineering Conference and
Symposium on the Foundations of Software Engineering}, 2018,
pp.~908--911.

{[}35{]} D. R. Cox, ``Regression models and life-tables,'' \emph{Journal
of the Royal Statistical Society: Series B (Methodological)}, vol.~34,
no. 2, pp.~187--202, 1972.

{[}36{]} F. E. Harrell, Jr., R. M. Califf, D. B. Pryor, K. L. Lee, and
R. A. Rosati, ``Evaluating the yield of medical tests,'' \emph{JAMA},
vol.~247, no. 18, pp.~2543--2546, 1982.

{[}37{]} C. Davidson-Pilon, ``lifelines: Survival analysis in Python,''
\emph{Journal of Open Source Software}, vol.~4, no. 40, p.~1317, 2019.

{[}38{]} T. M. Therneau and P. M. Grambsch, \emph{Modeling Survival
Data: Extending the Cox Model}. New York: Springer, 2000.

{[}39{]} H. Putter, M. Fiocco, and R. B. Geskus, ``Tutorial in
biostatistics: Competing risks and multi-state models,''
\emph{Statistics in Medicine}, vol.~26, no. 11, pp.~2389--2430, 2007.

{[}40{]} T. Chen and C. Guestrin, ``XGBoost: A scalable tree boosting
system,'' in \emph{Proceedings of the 22nd ACM SIGKDD International
Conference on Knowledge Discovery and Data Mining}, 2016, pp.~785--794.

{[}41{]} K. El Emam, S. Benlarbi, N. Goel, and S. N. Rai, ``The
confounding effect of class size on the validity of object-oriented
metrics,'' \emph{IEEE Transactions on Software Engineering}, vol.~27,
no. 7, pp.~630--650, 2001.

{[}42{]} J. P. Fine and R. J. Gray, ``A proportional hazards model for
the subdistribution of a competing risk,'' \emph{Journal of the American
Statistical Association}, vol.~94, no. 446, pp.~496--509, 1999.

{[}43{]} S. Suissa, ``Immortal time bias in pharmacoepidemiology,''
\emph{American Journal of Epidemiology}, vol.~167, no. 4, pp.~492--499,
2008.

{[}44{]} D. Schoenfeld, ``Partial residuals for the proportional hazards
regression model,'' \emph{Biometrika}, vol.~69, no. 1, pp.~239--241,
1982.

{[}45{]} D. G. Clayton, ``A model for association in bivariate life
tables and its application in epidemiological studies of familial
tendency in chronic disease incidence,'' \emph{Biometrika}, vol.~65, no.
1, pp.~141--151, 1978.

{[}46{]} M. H. Gail, S. Wieand, and S. Piantadosi, ``Biased estimates of
treatment effect in randomized experiments with nonlinear regressions
and omitted covariates,'' \emph{Biometrika}, vol.~71, no. 3,
pp.~431--444, 1984.

{[}47{]} D. Y. Lin and L. J. Wei, ``The robust inference for the Cox
proportional hazards model,'' \emph{Journal of the American Statistical
Association}, vol.~84, no. 408, pp.~1074--1078, 1989.

{[}48{]} Y. Wang, H. Le, A. D. Gotmare, N. D. Q. Bui, J. Li, and S. C.
H. Hoi, ``CodeT5+: Open code large language models for code
understanding and generation,'' in \emph{Proceedings of the 2023
Conference on Empirical Methods in Natural Language Processing}, 2023,
pp.~1069--1088.

{[}49{]} M. Kim, V. Sazawal, D. Notkin, and G. Murphy, ``An empirical
study of code clone genealogies,'' in \emph{Proceedings of the 10th
European Software Engineering Conference held jointly with the 13th ACM
SIGSOFT International Symposium on Foundations of Software Engineering},
2005, pp.~187--196.

{[}50{]} E. da S. Maldonado, R. Abdalkareem, E. Shihab, and A.
Serebrenik, ``An empirical study on the removal of self-admitted
technical debt,'' in \emph{Proceedings of the IEEE International
Conference on Software Maintenance and Evolution}, 2017, pp.~238--248.

{[}51{]} F. Zampetti, A. Serebrenik, and M. Di Penta, ``Was
self-admitted technical debt removal a real removal?'' in
\emph{Proceedings of the 15th International Conference on Mining
Software Repositories}, 2018, pp.~526--536.

{[}52{]} D. Caivano, P. Cassieri, S. Romano, and G. Scanniello, ``On the
spread and evolution of dead methods in Java desktop applications: an
exploratory study,'' \emph{Empirical Software Engineering}, vol.~28, no.
3, p.~64, 2023.

{[}53{]} S. Habchi, R. Rouvoy, and N. Moha, ``On the survival of Android
code smells in the wild,'' in \emph{Proceedings of the IEEE/ACM 6th
International Conference on Mobile Software Engineering and Systems},
2019, pp.~87--98.

{[}54{]} Anthropic, ``Claude Opus 5 and Claude Opus 5.5,'' large
language models, 2026. {[}Online{]}. Available:
https://www.anthropic.com/claude

\end{document}